\documentclass[12pt]{article}
\usepackage{enumerate}
\usepackage{amsfonts}
\usepackage{amsmath}
\usepackage{amssymb}
\usepackage{amsthm}

\def\H{\mathcal{H}}

\def\S{\mathfrak{S}}

\def\B{\mathfrak{B}}

\newcommand{\id}{\mathrm{Id}}
\newcommand{\Tr}{\mathrm{Tr}}

\newcounter{defin}  \newcounter{lemma}  \newcounter{theorem}
\newcounter{property} \newcounter{corol}  \newcounter{remark} \newcounter{example}

\newenvironment{lemma}{\par\refstepcounter{lemma}
     \textbf{Lemma \thelemma.} }{\rm\par}
\newenvironment{theorem}{\par\refstepcounter{theorem}
     \textbf{Theorem \thetheorem.}\ }{\rm\par}

\newenvironment{corollary}{\par\refstepcounter{corol}
     \textbf{Corollary \thecorol.} }{\rm\par}
\newenvironment{definition}{\par\refstepcounter{defin}
     \textbf{Definition \thedefin.}\ }{\rm\par}
\newenvironment{remark}{\par\refstepcounter{remark}
     \textbf{Remark \theremark.}}{\rm\par}

\begin{document}

\title{On channels with positive quantum zero-error capacity having vanishing $n$-shot capacity}

\author{M.E. Shirokov\footnote{Steklov Mathematical Institute, RAS, Moscow, email:msh@mi.ras.ru}}
\date{}
\maketitle

\begin{abstract}
We show that unbounded number of channel uses may be necessary for
perfect transmission of  quantum information. For any $n$ we
explicitly construct low-dimensional quantum channels (input
dimension  4, Choi rank  2 or 4) whose quantum zero-error capacity
is positive but the corresponding $n$-shot capacity is zero. We give
estimates for quantum zero-error capacity of such channels (as a
function of $n$) and show that these channels can be chosen in any
small vicinity (in the $cb$-norm) of a classical-quantum channel.

Mathematically, this property means appearance of an ideal
(noiseless) subchannel only in sufficiently large tensor power of a
channel.

Our approach (using special continuous deformation of a maximal
commutative $*$-subalgebra of $M_4$) also gives  low-dimensional
examples of superactivation of 1-shot quantum zero-error capacity.

Finally, we consider multi-dimensional construction which gives
channels with greater values of quantum zero-error capacity and
vanishing $n$-shot capacity.
\end{abstract}
\maketitle

\section{Introduction}

It is well known that the rate of information transmission
over classical and quantum communication channels can be increased
by simultaneous use of many copies of a channel. It is this fact
that implies necessity of regularization in definitions of different
capacities of a channel \cite{H-SCI,N&Ch}.\smallskip

In this paper we show that perfect transmission of quantum
information over a quantum channel may require unbounded number of
channel uses. We prove by explicit construction that for any given
$n$ there is a channel $\Phi_n$ such that
\begin{equation}\label{one-e}
\bar{Q}_0(\Phi_n^{\otimes n})=0,\quad \textrm{but} \quad
Q_0(\Phi_n)>0,
\end{equation}
where $\bar{Q}_0$ and $Q_0$ are respectively the 1-shot and the
asymptotic quantum zero-error capacities defined in Section 2.
\smallskip

This effect is closely related to the recently discovered phenomenon
of superactivation of zero-error capacities \cite{CCH,C&S,Duan}.
Indeed, (\ref{one-e}) is equivalent to existence of $m>n$ such that
\begin{equation}\label{one-f}
\bar{Q}_0(\Phi_n)=\bar{Q}_0(\Phi_n^{\otimes2})=\ldots=\bar{Q}_0(\Phi_n^{\otimes
(m-1)})=0,\quad \textrm{but} \quad \bar{Q}_0(\Phi_n^{\otimes m})>0.
\end{equation}
Mathematically, (\ref{one-f}) means that all the channels $\Phi_n,
\Phi_n^{\otimes2},\ldots,\Phi_n^{\otimes (m-1)}$ have no ideal
(noiseless) subchannels but the channel $\Phi_n^{\otimes m}$
has.\smallskip

We show how for any given $n$ to explicitly construct a
pseudo-diagonal quantum channel $\Phi_n$ with the input dimension
$d_A=4$ and the Choi rank $d_E\geq2$ satisfying (\ref{one-f}) by
determining its noncommutative graph. We also obtain the estimate
for $m$ as a function of $n$, which gives the lower bound for
$Q_0(\Phi_n)$ in (\ref{one-e}). This shows that
\begin{equation}\label{one-g}
\sup_{\Phi}\left\{Q_0(\Phi)\,|\,\bar{Q}_0(\Phi^{\otimes
n})=0\,\right\}\geq \frac{2\ln(3/2)}{\pi n}\qquad \forall n.
\end{equation}
It is also observed that a channel $\Phi_n$ satisfying (\ref{one-e})
and (\ref{one-f}) can be obtained by arbitrarily small deformation (in
the $cb$-norm) of a classical-quantum channel with
$d_A=d_E=4$.\smallskip

The main problem in finding the channel $\Phi_n$ is to show
nonexistence of error correcting codes for the channel
$\Phi_n^{\otimes n}$ (provided the existence of such codes is proved
for $\Phi_n^{\otimes m}$). We solve this problem by using the
special continuous deformation of a maximal commutative $*$-subalgebra
of $4\times4$ matrices as the noncommutative graph of $\Phi_n$ and by
noting that the Knill-Laflamme error-correcting conditions are violated
for any maximal commutative $*$-subalgebra with the positive
dimension-independent gap (Lemma \ref{b-lemma}).

Our construction also gives  low-dimensional examples of
superactivation of 1-shot quantum zero-error capacity. In
particular, it gives an example of symmetric superactivation with
$d_A=4, d_E=2$ (simplifying the example in \cite{Sh&Sh+}) and shows
that such superactivation is possible for two channels with
$\,d_A=d_E=4\,$ if one of them is arbitrarily close (in the
$cb$-norm) to a classical-quantum channel.

In the last section we consider multi-dimensional generalization of
our basic construction. It gives examples of channels which amplify
the lower bound in (\ref{one-g}) by the factor
$\frac{2\ln2}{\pi}\approx2.26$. Unfortunately, we did not managed
to show that the value in the left side of (\ref{one-g}) is
$+\infty$ (as it is reasonable to conjecture). Estimation of this
value remains an open question.

\section{Preliminaries}

Let
$\Phi:\mathfrak{S}(\mathcal{H}_A)\rightarrow\mathfrak{S}(\mathcal{H}_B)$
be a quantum channel, i.e. a  completely positive trace-preserving
linear map \cite{H-SCI,N&Ch}. Stinespring's theorem implies the
existence of a Hilbert space $\mathcal{H}_E$ and of an isometry
$V:\mathcal{H}_A\rightarrow\mathcal{H}_B\otimes\mathcal{H}_E$ such
that
\begin{equation}\label{Stinespring-rep}
\Phi(\rho)=\mathrm{Tr}_{\mathcal{H}_E}V\rho V^{*},\quad
\rho\in\mathfrak{S}(\mathcal{H}_A).
\end{equation}
The minimal dimension of $\mathcal{H}_E$ is called Choi rank of
$\Phi$ and denoted $d_E$.

The quantum  channel
\begin{equation}\label{c-channel}
\mathfrak{S}(\mathcal{H}_A)\ni
\rho\mapsto\widehat{\Phi}(\rho)=\mathrm{Tr}_{\mathcal{H}_B}V\rho
V^{*}\in\mathfrak{S}(\mathcal{H}_E)
\end{equation}
is called \emph{complementary} to the channel $\Phi$
\cite{H-SCI,H-c-c}. The complementary channel is defined uniquely up
to isometrical equivalence \cite[the Appendix]{H-c-c}.

The 1-shot  quantum zero-error capacity $\bar{Q}_0(\Phi)$ of a
channel $\Phi$ is defined as $\;\sup_{\H\in
q_0(\Phi)}\log_2\dim\H\,$, where $q_0(\Phi)$ is the set of all
subspaces $\H_0$ of $\H_A$ on which the channel $\Phi$ is perfectly
reversible (in the sense that there is a channel $\Theta$ such that
$\Theta(\Phi(\rho))=\rho$ for all states $\rho$ supported by
$\H_0$). Any subspace $\H_0\in q_0(\Phi)$ is called \emph{error
correcting code} for the channel $\Phi$ \cite{W&Co, H-SCI}.

The (asymptotic)  quantum zero-error capacity is defined by
regularization: $Q_0(\Phi)=\sup_n n^{-1}\bar{Q}_0(\Phi^{\otimes n})$
\cite{Duan, W&Co}.

It is well known that a channel $\Phi$ is perfectly reversible on a
subspace $\H_0$ if and only if the restriction of the complementary
channel $\widehat{\Phi}$ to the subset $\S(\H_0)$ is completely
depolarizing , i.e. $\widehat{\Phi}(\rho_1)=\widehat{\Phi}(\rho_2)$
for all states $\rho_1$ and $\rho_2$ supported by $\H_0$
\cite[Ch.10]{H-SCI}. It follows that the 1-shot quantum zero-error
capacity $\bar{Q}_0(\Phi)$ of a channel $\Phi$ is completely
determined by the set
$\mathcal{G}(\Phi)\doteq\widehat{\Phi}^*(\B(\H_E))$ called the
\emph{noncommutative graph} of $\Phi$ \cite{W&Co}. In particular,
the Knill-Laflamme error-correcting condition \cite{K-L} implies the
following lemma.\smallskip

\begin{lemma}\label{trans-l+}
\emph{A set $\{\varphi_k\}_{k=1}^d$ of unit orthogonal vectors in
$\H_A$ is a basis of error-correcting code for a channel
$\,\Phi:\S(\H_A)\rightarrow\S(\H_B)$ if and only if
\begin{equation}\label{operators+}
\langle \varphi_l|A|\varphi_k\rangle=0\quad\textit{and}\quad \langle
\varphi_l|A|\varphi_l\rangle=\langle
\varphi_k|A|\varphi_k\rangle\quad\forall
A\in\mathfrak{L},\;\,\forall k\neq l,
\end{equation}
where $\mathfrak{L}$ is any subset of $\,\B(\H_A)$ such that
$\,\mathrm{lin}\mathfrak{L}=\mathcal{G}(\Phi)$.}
\end{lemma}\smallskip

This lemma shows that $\,\bar{Q}_0(\Phi)\geq\log_2 d\,$ if and only
if there exists a set $\{\varphi_k\}_{k=1}^d$ of unit vectors in
$\H_A$ satisfying condition (\ref{operators+}).\smallskip

\begin{remark}\label{trans-l+r}
Since a subspace $\mathfrak{L}$ of the algebra $\mathfrak{M}_n$ of
$n\times n$ matrices is a noncommutative graph of a particular
channel if and only if
\begin{equation}\label{L-cond}
\mathfrak{L}\;\,\textup{is
symmetric}\;\,(\mathfrak{L}=\mathfrak{L}^*)\;\,\textup{and contains
the unit matrix}\;I_n
\end{equation}
(see Lemma 2 in \cite{Duan} or Proposition 2 in \cite{Sh&Sh}), Lemma
\ref{trans-l+} shows that one can "construct" a channel $\Phi$ with
$\dim\H_A=n$ having positive (correspondingly, zero) 1-shot quantum
zero-error  capacity by taking a subspace
$\mathfrak{L}\subset\mathfrak{M}_n$ satisfying (\ref{L-cond}) for
which the following condition is valid (correspondingly, not valid)
\begin{equation}\label{l-3-c}
\exists\varphi,\psi\in[\mathbb{C}^n]_1 \;\;\textup{s.t.}\;\; \langle
\psi|A|\varphi\rangle=0\;\;\textup{and}\;\; \langle
\varphi|A|\varphi\rangle=\langle \psi|A|\psi\rangle\quad\forall
A\in\mathfrak{L},
\end{equation}
where $[\mathbb{C}^n]_1$ is the unit sphere of $\mathbb{C}^n$.
$\square$
\end{remark}\medskip

We will use the following two notions.\smallskip
\begin{definition}\label{c-q-def} \cite{H-SCI}
A finite-dimensional\footnote{In infinite dimensions there exist
channels naturally called classical-quantum, which have no
representation   (\ref{c-q-rep}).} channel
$\Phi:\S(\H_A)\rightarrow\S(\H_B)$ is called
\emph{classical-quantum} if it has the representation
\begin{equation}\label{c-q-rep}
\Phi(\rho)=\sum_{k}\langle k|\rho|k\rangle\sigma_k,
\end{equation}
where $\{|k\rangle\}$ is an orthonormal basis in $\H_A$ and
$\{\sigma_k\}$ is a collection of states in $\S(\H_B)$.
\end{definition}\medskip

\begin{definition}\label{p-d-def} \cite{R} A finite-dimensional channel $\Phi:\S(\H_A)\rightarrow\S(\H_B)$ is called
\emph{pseudo-diagonal} if it has the representation
$$
\Phi(\rho)=\sum_{i,j}c_{ij}\langle
\psi_i|\rho|\psi_j\rangle|i\rangle\langle j|,
$$
where $\{c_{ij}\}$ is a Gram matrix of a collection of unit vectors,
$\{|\psi_i\rangle\}$ is a collection of vectors in $\H_A$ such that
$\;\sum_i |\psi_i\rangle\langle \psi_i|=I_{\H_A}\,$  and
$\{|i\rangle\}$ is an orthonormal basis in $\H_B$.
\end{definition}

Pseudo-diagonal channels are complementary to entanglement-breaking
channels and vice versa \cite{R,H-c-c}.

For any matrix $A\in\mathfrak{M}_n$ denote by
$\mathrm{\Upsilon}_{A}$ the operator of Schur multiplication by $A$
in $\mathfrak{M}_n$ (also called the Hadamard multiplication). Its
$cb$-norm will be denoted $\|\mathrm{\Upsilon}_{A}\|_{\mathrm{cb}}$.
It coincides with the operator norm of $\mathrm{\Upsilon}_{A}$ and
is also called the Schur (or Hadamard) multiplier norm of $A$
\cite{M,P}.

\section{Basic example}

For any given $\,\theta\in\mathrm{T}\doteq(-\pi,\pi]\,$ consider the
subspace
\begin{equation}\label{L-def+}
\mathfrak{L}_{\theta} = \left\{M=\left[\begin{array}{cccc}
a &  b & \gamma c &  d\\
b &  a & d &  \bar{\gamma} c\\
\bar{\gamma}c &  d & a &  b\\
d &  \gamma c & b &  a
\end{array}\right]\!,\; a,b,c,d\in\mathbb{C},\; \gamma=\exp\left(\textstyle\frac{\mathrm{i}}{2}\theta\right)\right\}
\end{equation}
of $\mathfrak{M}_4$. This subspace satisfies condition
(\ref{L-cond}) and has the following property
\begin{equation}\label{W-inv}
    A=W_4^*AW_4\quad \forall A\in \mathfrak{L}_{\theta},\quad \textrm{where}\quad W_4=\left[\begin{array}{cccc}
0 &  0 & 0 &  1\\
0 &  0 & 1 &  0\\
0 &  1 & 0 &  0\\
1 &  0 & 0 &  0
\end{array}\right].
\end{equation}

Denote by $\widehat{\mathfrak{L}}_{\theta}$ the set of all channels
whose noncommutative graph coincides with $\mathfrak{L}_{\theta}$.
For each $\theta$ the set $\widehat{\mathfrak{L}}_{\theta}$ contains
infinitely many different channels with $d_A\doteq\dim\H_A=4$ and
$d_E\geq2$.\smallskip

\begin{lemma}\label{cfc}
1) \emph{There is a  family $\,\{\Phi^1_{\theta}\}$ of
pseudo-diagonal channels (see Def.\ref{p-d-def}) with $d_E=2$ such
that $\,\Phi^1_{\theta}\in\widehat{\mathfrak{L}}_{\theta}$ for each
$\theta$.}

2) \emph{There is a family $\,\{\Phi^2_{\theta}\}$ of
pseudo-diagonal channels with $d_E=4$ such that
$\,\Phi^2_{\theta}\in\widehat{\mathfrak{L}}_{\theta}$ for each
$\theta$ and $\,\Phi^2_{0}$ is a classical-quantum channel (see
Def.\ref{c-q-def}).}

\emph{The  families $\,\{\Phi^1_{\theta}\}$ and
$\,\{\Phi^2_{\theta}\}$ can be chosen continuous in the following
sense:
\begin{equation}\label{c-rep}
\Phi^k_{\theta}(\rho)=\mathrm{Tr}_{\mathcal{H}^k_E}V^k_{\theta}\rho
[V^k_{\theta}]^{*},\quad \rho\in\mathfrak{S}(\mathcal{H}_A),\quad
k=1,2,
\end{equation}
where $\,V^1_{\theta}$, $V^2_{\theta}$ are continuous families of
isometries,  $\mathcal{H}^1_E=\mathbb{C}^2$,
$\mathcal{H}^2_E=\mathbb{C}^4$.}\footnote{This implies continuity of
these families in the $cb$-norm \cite{K&Co}.}
\end{lemma}\smallskip

Lemma \ref{cfc} is proved in the Appendix by explicit construction
of representations (\ref{c-rep}).
\medskip 

\begin{theorem}\label{sqc}  \emph{Let  $\,\Phi_{\theta}$ be a  channel in $\,\widehat{\mathfrak{L}}_{\theta}$ and $n\in\mathbb{N}$ be arbitrary.}\medskip

A) \emph{$\bar{Q}_0(\Phi_{\theta})>0\;$  if and only if
$\;\theta=\pi\,$ and $\;\bar{Q}_0(\Phi_{\pi})=1$.}

\medskip

B) \emph{If $\;\theta_1+\ldots+\theta_n=\pi(\mathrm{mod}\,2\pi)\,$
then
$\;\bar{Q}_0(\Phi_{\theta_1}\otimes\ldots\otimes\Phi_{\theta_n})>0\;$
and there exist $\,2^n$ mutually orthogonal $\,2\textup{-}D$ error
correcting codes for the channel
$\,\Phi_{\theta_1}\otimes\ldots\otimes\Phi_{\theta_n}$. For each
binary $n$-tuple $(x_1,\ldots x_n)$ the corresponding error
correcting code is spanned by the images of the vectors
\begin{equation}\label{main-vec}
|\varphi\rangle=\textstyle{\frac{1}{\sqrt{2}}}\left[\;|1\ldots
1\rangle+\mathrm{i}\,|2\ldots 2\rangle\,\right],\quad
|\psi\rangle=\textstyle{\frac{1}{\sqrt{2}}}\left[\;|3\ldots
3\rangle+\mathrm{i}\,|4\ldots 4\rangle\,\right]\!,
\end{equation}
under the unitary transformation $\,U_{x_1}\otimes\ldots\otimes
U_{x_n}$, where $\{|1\rangle, \ldots, |4\rangle\}$ is the canonical
basis in $\mathbb{C}^4$, $\,U_0=I_4$ and $\,U_1=W_4$ (defined in
(\ref{W-inv})).}
\medskip

C) \emph{If $\;|\theta_1|+\ldots+|\theta_n|\leq 2\ln(3/2)\,$ then
$\;\bar{Q}_0(\Phi_{\theta_1}\otimes\ldots\otimes\Phi_{\theta_n})=0\;$.}
\end{theorem}
\medskip

\medskip
\begin{remark}\label{sqc-r} It is easy to show that
$\;\bar{Q}_0(\Phi^{\otimes n}_{\theta})=\bar{Q}_0(\Phi^{\otimes
n}_{-\theta})\;$ and that the set of all $\theta$ such that
$\bar{Q}_0(\Phi^{\otimes n}_{\theta})=0$ is open. Hence for each $n$
there is $\varepsilon_n>0\,$ such that $\;\bar{Q}_{0}(\Phi^{\otimes
n}_{\theta})=0\,$ if $\;|\theta|<\varepsilon_n$ and
$\;\bar{Q}_{0}(\Phi^{\otimes n}_{\pm\varepsilon_n})>0\,$. Theorem
\ref{sqc} shows that $\,\varepsilon_1=\pi$ and
$\,2\ln(3/2)/n<\varepsilon_n\leq\pi/n\,$ for $\,n>1$. Since
assertion C is proved by using quite coarse estimates, one can
conjecture that $\,\varepsilon_n=\pi/n\,$ for $n>1$. There exist
some arguments confirming validity of this conjecture for $\,n=2$.
\end{remark}
\smallskip

\begin{remark}\label{sqc-rr} Assertion B of Theorem 1 can be
strengthened as follows:\smallskip

B') \emph{If $\;\theta_1+\ldots+\theta_n=\pi(\mathrm{mod}\,2\pi)\,$
then there exist  $\,2^n$ mutually orthogonal $\,2\textup{-}D$
projectors $P_{\bar{x}}$ indexed by a binary $n$-tuple
$\,\bar{x}=(x_1,\ldots x_n)$ such that
$$
P_{\bar{x}}AP_{\bar{x}}=\lambda(A)P_{\bar{x}}\quad \forall
A\in\mathfrak{L}_{\theta_1}\otimes\ldots\otimes\mathfrak{L}_{\theta_n},
$$
where $\lambda(A)\in\mathbb{C}$ does not depend on $\,\bar{x}$.
$P_{\bar{x}}$ is the projector on the subspace
$\,U_{x_1}\otimes\ldots\otimes U_{x_n}(\H_0)$, where $\H_0$ is the
linear hull of vectors (\ref{main-vec}).}\smallskip

So, in the orthonormal basis $\,\left\{U_{x_1}\otimes\ldots\otimes
U_{x_n}|\varphi\rangle, U_{x_1}\otimes\ldots\otimes
U_{x_n}|\psi\rangle, \ldots \right\}$ the main
$\,2^{n+1}\times2^{n+1}$ minor of all matrices in
$\mathfrak{L}_{\theta_1}\otimes\ldots\otimes\mathfrak{L}_{\theta_n}$
has the form
\begin{equation}\label{min-f}
\left[\begin{array}{cccc}
\lambda I_2 &  * & \cdots &  * \\
* &  \lambda I_2 & \cdots &  * \\
\cdots &  \cdots & \cdots &  *  \\
*      &    *    &  *     &  \lambda I_2
 \end{array}\right]\! , \;\; \textrm{where}\;
\lambda\in\mathbb{C},\;\, I_2\;\, \textrm{is the unit}\;\,
2\times2\;\, \textrm{matrix}.
\end{equation}
\end{remark}

Theorem 1 implies the main result of this paper.\smallskip

\begin{corollary}\label{sqc-c+} \emph{Let $\;n\,$ be arbitrary
and $\;m\,$ be a natural number such that $\theta_*=\pi/m\leq
2\ln(3/2)/n$. Then}
\begin{equation}\label{main-r}
\bar{Q}_0(\Phi_{\theta_*}^{\otimes n})=0\;\;\;
\textit{but}\;\;\;\bar{Q}_0(\Phi_{\theta_*}^{\otimes
m})\geq1\;\;\;\textit{and hence}\;\;\; Q_0(\Phi_{\theta_*})\geq 1/m.
\end{equation}
\emph{There exist $\,2^m$ mutually orthogonal $\,2$-D error
correcting codes for the channel $\,\Phi_{\theta_*}^{\otimes m}$.}
\end{corollary}
\medskip

Relation (\ref{main-r}) means that it is not possible to transmit
any quantum information with no errors by using  $\leq n$ copies of
the channel $\Phi_{\theta_*}$, but such  transmission is possible if
the number of copies is $\geq m$.
\medskip

\begin{remark}\label{sqc-c+r}
In (\ref{main-r}) one can take $\Phi_{\theta_*}=\Phi^1_{\theta_*}$
-- a channel from the family described in the first part of Lemma
\ref{cfc}. So,  Corollary \ref{sqc-c+} shows that for any $n$ there
exists a channel $\Phi_n$ with $d_A=4$ and $d_E=2$ such that
$\bar{Q}_0(\Phi_n^{\otimes n})=0$ and
$$
Q_0(\Phi_n)\geq \left(\left[\frac{\pi
n}{2\ln(3/2)}\right]+1\right)^{-1}=\frac{2\ln(3/2)}{\pi
n}+o(1/n),\quad n\rightarrow+\infty,
$$
where $[x]$ is the integer part of $x$.
\end{remark}\medskip

It is natural to ask about the maximal value of quantum zero-error
capacity of a channel with given input dimension having vanishing
$n$-shot capacity, i.e. about the value
\begin{equation}\label{s-d-def}
S_d(n)\doteq\sup_{\Phi\,:\,d_A=d}\left\{Q_0(\Phi)\,|\,\bar{Q}_0(\Phi^{\otimes
n})=0\,\right\},
\end{equation}
where the supremum is over all quantum channels with
$d_A\doteq\dim\H_A=d$. We may also consider the value
\begin{equation}\label{s-star-def}
S_*(n)\doteq\sup_{d}S_d(n)=\lim_{d\rightarrow+\infty}S_d(n)\leq+\infty.
\end{equation}
The sequences $\{S_d(n)\}_n$ and $\{S_*(n)\}_n$ are non-increasing
and the first of them is bounded by $\log_2 d$. Theorem 2 in
\cite{Sh&Sh+} shows that
$$
S_{2d}(1)\geq \frac{\log_2 d}{2}\quad\textrm{and hence}\quad
S_{*}(1)=+\infty.
$$
It seems reasonable to conjecture that $S_*(n)=+\infty$ for all $n$.
A possible way to prove this conjecture is discussed at the end of
Section 4.\medskip

It follows from the superadditivity of quantum zero-error capacity
that
\begin{equation}\label{s-rel}
S_{d^k}(n)\geq kS_{d}(nk)\quad\textrm{and hence}\quad S_{*}(n)\geq
kS_{*}(nk)\quad\textrm{for any}\;\,k,n.
\end{equation}
These relations show that the assumption $\,S_*(n_0)<+\infty\,$ for
some $\,n_0$ implies
$$
S_d(n)= O(1/n)\;\, \textrm{for each}\;\, d \quad\textrm{and}\quad
S_*(n)= O(1/n)\;\,\textrm{if}\;\, n\geq n_0.
$$

By Corollary \ref{sqc-c+} we have
\begin{equation}\label{s-e-0}
S_{4}(n)\geq \left(\left[\frac{\pi
n}{2\ln(3/2)}\right]+1\right)^{-1}=\frac{2\ln(3/2)}{\pi
n}+o(1/n),\qquad \forall n.
\end{equation}
This and (\ref{s-rel}) imply the estimation
\begin{equation}\label{s-e-1}
S_{4^k}(n)\geq k\,\frac{2\ln(3/2)}{\pi
kn}+o(1/(kn))=\frac{2\ln(3/2)}{\pi n}+o(1/(kn)),
\end{equation}
which shows that
\begin{equation}\label{s-e-2}
S_*(n)\geq\frac{2\ln(3/2)}{\pi n}\qquad \forall n.
\end{equation}
In Section 4 we will improve these  lower bounds by considering the
multi-dimensional generalization of the above construction.\medskip

\begin{remark}\label{sa-nr} Since the parameter $\,\theta_*$ in Corollary \ref{sqc-c+} can be
taken arbitrarily close to zero, the second part of Lemma \ref{cfc}
shows that the channel $\Phi_{\theta_*}$, for which
$\,\bar{Q}_0(\Phi^{\otimes n}_{\theta_*})=0\,$ and
$\,Q_0(\Phi_{\theta_*})>0\,$, can be chosen in any small vicinity
(in the $cb$-norm) of the classical-quantum channel $\Phi^2_0$.
\end{remark}\medskip

Theorem \ref{sqc} also gives examples of superactivation of 1-shot
quantum zero-error  capacity.\smallskip

\begin{corollary}\label{sa-c} \emph{If $\,\theta\neq0,\pi$ then  the following
superactivation property
$$
\,\bar{Q}_0(\Phi_{\theta})=\bar{Q}_0(\Phi_{\pi-\theta})=0\quad and
\quad \bar{Q}_0(\Phi_{\theta}\otimes\Phi_{\pi-\theta})>0
$$
holds for any channels
$\,\Phi_{\theta}\in\widehat{\mathfrak{L}}_{\theta}$ and
$\,\Phi_{\pi-\theta}\in\widehat{\mathfrak{L}}_{\pi-\theta}$. For any
$\;\theta\in\mathrm{T}\,$ there exist $4$ mutually orthogonal
$\,2\textrm{-}D$ error correcting codes for the channel
$\,\Phi_{\theta}\otimes\Phi_{\pi-\theta}$, one of them is spanned by
the vectors
\begin{equation}\label{sp-vec}
|\varphi\rangle=\textstyle{\frac{1}{\sqrt{2}}}\left[\;|11\rangle+\mathrm{i}\,|22\rangle\,\right],\quad
|\psi\rangle=\textstyle{\frac{1}{\sqrt{2}}}\left[\;|3
3\rangle+\mathrm{i}\,|44\rangle\,\right],
\end{equation}
others are the images of this subspace under the unitary
transformations $I_4\otimes W_4$, $W_4\otimes I_4$ and $\,W_4\otimes
W_4$ (the operator $W_4$ is defined in (\ref{W-inv})).}
\end{corollary}
\medskip

\begin{remark}\label{sa-r} Corollary \ref{sa-c} shows that the channel
$\Phi^1_{\pi/2}$ (taken from the fist part of Lemma \ref{cfc}) is an
example of symmetric superactivation of 1-shot quantum zero-error
capacity \emph{with Choi rank $2$}.\footnote{This strengthens the
result in \cite{Sh&Sh+}, where a similar example with Choi rank $3$
and the same input dimension was constructed.}\smallskip

By taking the family $\{\Phi^2_{\theta}\}$ from the second part of
Lemma \ref{cfc} and tending $\,\theta\,$ to zero we see from
Corollary \ref{sa-c} that \emph{the superactivation of 1-shot
quantum zero-error capacity may hold for two channels with
$d_A=d_E=4$ if one of them is arbitrarily close (in the $cb$-norm)
to a classical-quantum channel}.
\smallskip

Note that the entangled subspace spanned by the vectors
(\ref{sp-vec}) is an error correcting code for the channel
$\,\Phi^2_{0}\otimes\Phi^2_{\pi}$ (and hence for the channel
$\,\Phi^2_{0}\otimes\id_{\mathbb{C}^4}$) despite the fact that
$\Phi^2_{0}$ is a classical-quantum channel.
\end{remark}

\medskip

\emph{Proof of Theorem 1.} A) It is easy to verify that the subspace
$\mathfrak{L}_{\pi}$ satisfies condition (\ref{l-3-c}) with the
vectors
$|\varphi\rangle=[1,\mathrm{i},0,0]^{\top},\,|\psi\rangle=[0,0,1,\mathrm{i}]^{\top}$.
\smallskip

To show that $\bar{Q}_0(\Phi_{\theta})=0$ for all $\theta\neq \pi$
represent the matrix $M$ in (\ref{L-def+}) as $M=A+cB$, where
$$
A=\left[\begin{array}{cccc}
a &  b & c &  d\\
b &  a & d &  c\\
c &  d & a &  b\\
d &  c & b &  a
\end{array}\right],\qquad B=\left[\begin{array}{cccc}
0 &  0 & \tau &  0\\
0 &  0 & 0 &  \bar{\tau}\\
\bar{\tau} &  0 & 0 &  0\\
0 &  \tau  & 0 &  0
\end{array}\right],\quad \tau=\gamma-1.
$$
Let $ \;S=\frac{1}{2}\left[\begin{array}{rrrr}
\!1 &  1 & 1 &  \;1\\
\!-1 &  1 & -1 & \;1\\
\!-1 &  -1 & 1 &  \;1\\
\!1 &  -1 & -1 &  \;1
\end{array}\right]\;
$ then $\;S^{-1}=S^{\top}=\frac{1}{2}\left[\begin{array}{rrrr}
1 &  -1 & -1 &  1\\
1 &  1 & -1 &  -1\\
1 &  -1 & 1 &  -1\\
1 &  1 & 1 &  1
\end{array}\right]\;
$ and
$$
S^{-1}AS=\left[\begin{array}{rrrr}
\tilde{a} &  0 & 0 &  0\\
0 &  \tilde{b} & 0 &  0\\
0 &  0 & \tilde{c} &  0\\
0 &  0 & 0 &  \tilde{d}
\end{array}\right],\quad S^{-1}BS=\textstyle\left[\begin{array}{rrrr}
u &  0 & 0 &  v\\
0 &  u & v &  0\\
0 &  -v & -u &  0\\
-v &  0 & 0 &  -u
\end{array}\right],
$$
where
$$
\begin{array}{l}
\tilde{a}=a-b-c+d,\qquad \tilde{b}=a+b-c-d,\qquad u=-\Re\tau=1-\Re\gamma\\
\tilde{c}=a-b+c-d,\qquad \tilde{d}=a+b+c+d, \qquad
v=\mathrm{i}\Im\tau=\mathrm{i}\Im\gamma.
\end{array}
$$
Thus the subspace $\mathfrak{L}_{\theta}$ is unitary equivalent to
the subspace
\begin{equation}\label{L-copy}
\mathfrak{L}^s_{\theta}=\left\{M=\left[\begin{array}{rrrr}
a &  0 & 0 &  0\\
0 &  b & 0 &  0\\
0 &  0 & c &  0\\
0 &  0 & 0 &  d
\end{array}\right]+\textstyle\frac{1}{4}(d+c-b-a)\,T_{\theta},\quad a,b,c,d\in\mathbb{C}\;\right\}
\end{equation}
where $T_{\theta}=S^{-1}BS\,$ is the above-defined matrix. Hence it
suffices to show that condition (\ref{l-3-c}) is not valid for
$\mathfrak{L}=\mathfrak{L}^s_{\theta}\,$ if $\,\theta\neq\pi$
(i.e.$\gamma\neq\mathrm{i}$).\smallskip

Assume the existence of unit vectors $|\varphi\rangle=[x_1, x_2,
x_3, x_4]^{\top}$ and $|\psi\rangle=[y_1, y_2, y_3, y_4]^{\top}$ in
$\mathbb{C}^4$ such that
\begin{equation}\label{l-3-c+}
\langle\psi | M | \varphi\rangle=0\;\; \textrm{and}\;\; \langle\psi
| M |\psi\rangle=\langle\varphi | M | \varphi\rangle\;\; \textrm{for
all}\;\; M\in\mathfrak{L}^s_{\theta}
\end{equation}
Since condition (\ref{l-3-c+}) is invariant under the rotation
$$
|\varphi\rangle\mapsto p|\varphi\rangle-q |\psi\rangle,
\quad|\psi\rangle\mapsto\bar{q}|\varphi\rangle+\bar{p}|\psi\rangle,\quad|p|^2+|q|^2=1,
$$
we may consider that $y_1=0$.

By taking successively $(a=-1, b=c=d=0)$, $(b=-1, a=c=d=0)$, $(c=1,
a=b=d=0)$ and $(d=1, a=b=c=0)$ we obtain from (\ref{l-3-c+}) the
following equations
$$
\bar{y}_1x_1=\bar{y}_2x_2=-\bar{y}_3x_3=-\bar{y}_4x_4=\textstyle\frac{1}{4}\langle\psi|T_{\theta}|\varphi\rangle,
$$
$$
|x_1|^2-|y_1|^2=|x_2|^2-|y_2|^2=|y_3|^2-|x_3|^2=|y_4|^2-|x_4|^2=
\textstyle\frac{1}{4}[\langle\varphi|T_{\theta}|\varphi\rangle-\langle\psi|T_{\theta}|\psi\rangle],
$$
Since $y_1=0$ and $\|\varphi\|=\|\psi\|=1$, the above equations
imply
\begin{equation*}
y_1=y_2=x_3=x_4=0
\end{equation*}
and
\begin{equation}\label{n-eq-2}
|x_1|^2=|x_2|^2=|y_3|^2=|y_4|^2=\textstyle\frac{1}{4}[\langle\varphi|T_{\theta}|\varphi\rangle-\langle\psi|T_{\theta}|\psi\rangle]=1/2.
\end{equation}
So, $|\varphi\rangle=[x_1, x_2, 0, 0\,]^{\top}$ and
$|\psi\rangle=[0, 0, y_3, y_4]^{\top}$, where $[x_1, x_2]^{\top}$
and $[y_3, y_4]^{\top}$ are unit vectors in $\mathbb{C}^2$ . It
follows from (\ref{n-eq-2}) that
$$
2=\left\langle\begin{array}{c}
\!x_1\! \\
\!x_2\!
\end{array}\right|\left.\begin{array}{rr}
\!\!u\! & \!0 \\
\!\!0\! & u
\end{array}\right|
\left.\begin{array}{c}
\!x_1\! \\
\!x_2\!
\end{array}\right\rangle-
\left\langle\begin{array}{c}
\!y_3\! \\
\!y_4\!
\end{array}\right|\left.\begin{array}{rr}
\!\!-u\! & \!\!0 \\
\!\!0\! & \!\!-u
\end{array}\right|
\left.\begin{array}{c}
\!y_3\! \\
\!y_4\!
\end{array}\right\rangle=2u,
$$
which can be valid only if $\gamma=\mathrm{i}$, i.e. $\theta=\pi$.

The above arguments also show that $\bar{Q}_0(\Phi_{\pi})=1$, since
the assumption $\bar{Q}_0(\Phi_{\pi})>1\,$ implies, by Lemma
\ref{trans-l+}, existence of orthogonal unit vectors $\,\phi_1$,
$\phi_2$, $\phi_3\,$ such that condition (\ref{l-3-c+}) with
$\varphi=\phi_i$, $\psi=\phi_j\,$ is valid for all $i\neq
j$.\smallskip

B) Let
$M_1\in\mathfrak{L}_{\theta_1},\ldots,M_n\in\mathfrak{L}_{\theta_n}$
be arbitrary and $X=M_1\otimes \ldots\otimes M_n$. To prove that the
linear hull $\H_0$ of  vectors (\ref{main-vec}) is an error
correcting code for the channel
$\,\mathrm{\Phi}_{\theta_1}\otimes\ldots\otimes\mathrm{\Phi}_{\theta_n}$
it suffices, by Lemma \ref{trans-l+}, to show that
\begin{equation}\label{one}
   \langle\psi | X | \varphi\rangle=0\quad\textrm{and}\quad
   \langle\psi | X |\psi\rangle=\langle\varphi | X | \varphi\rangle.
\end{equation}
We have
$$
\begin{array}{c}
2\langle\psi | X | \varphi\rangle= \langle 3\ldots 3| X|1\ldots
1\rangle+\mathrm{i}\langle 3\ldots 3| X|2\ldots
2\rangle-\mathrm{i}\langle 4\ldots 4| X|1\ldots 1\rangle\\\\+\langle
4\ldots 4|X|2\ldots 2\rangle= c_1 \ldots c_n(\bar{\gamma}_1
\ldots\bar{\gamma}_n+\gamma_1 \ldots\gamma_n)+d_1 \ldots
d_n(\mathrm{i}-\mathrm{i})=0,
\end{array}
$$
since $\gamma_1 \ldots\gamma_n=\pm\mathrm{i}$,
$$
\begin{array}{c}
2\langle\varphi | X | \varphi\rangle= \langle 1\ldots 1| X|1\ldots
1\rangle+\mathrm{i}\langle 1\ldots 1| X|2\ldots
2\rangle-\mathrm{i}\langle 2\ldots 2| X|1\ldots 1\rangle\\\\+\langle
2\ldots 2|X|2\ldots 2\rangle=a_1\ldots a_n(1+1)+b_1\ldots
b_n(\mathrm{i}-\mathrm{i})=2a_1\ldots a_n
\end{array}
$$
and
$$
\begin{array}{c}
2\langle\psi | X | \psi\rangle= \langle 3\ldots 3| X|3\ldots
3\rangle+\mathrm{i}\langle 3\ldots 3| X|4\ldots
4\rangle-\mathrm{i}\langle 4\ldots 4| X|3\ldots 3\rangle\\\\+\langle
4\ldots 4|X|4\ldots 4\rangle=a_1\ldots a_n(1+1)+b_1\ldots
b_n(\mathrm{i}-\mathrm{i})=2a_1\ldots a_n.
\end{array}
$$
Thus the both equalities in (\ref{one}) are valid.

To prove that the subspace $U_{\bar{x}}(\H_0)$, where
$\,U_{\bar{x}}=U_{x_1}\otimes\ldots\otimes U_{x_n}$, is an error
correcting code for the channel
$\,\mathrm{\Phi}_{\theta_1}\otimes\ldots\otimes\mathrm{\Phi}_{\theta_n}$
it suffices to note that (\ref{W-inv}) implies
$U^*_{\bar{x}}AU_{\bar{x}}=A$ for all
$A\in\mathfrak{L}_{\theta_1}\otimes\ldots\otimes\mathfrak{L}_{\theta_n}$.
\medskip

C) To show that
$\,\bar{Q}_0\left(\bigotimes_{k=1}^n\!\Phi_{\theta_k}\right)=0\,$ if
$\,\sum_{k=1}^n|\theta_k|\leq 2\ln(3/2)\,$  note that
$\mathfrak{L}_{\theta}=\Upsilon_{D(\theta)}(\mathfrak{L}_{0})$ and
$\bigotimes_{k=1}^n\mathfrak{L}_{\theta_k}=
\bigotimes_{k=1}^n\!\Upsilon_{D(\theta_k)}\left(\mathfrak{L}_{0}^{\otimes
n}\right)$, where $\Upsilon_{D(\theta)}$ is the Schur multiplication
by the matrix
\begin{equation}\label{tmp}
D(\theta)=\left[\begin{array}{cccc}
1 &  1 & \gamma &  1\\
1 &  1 & 1 &  \bar{\gamma}\\
\bar{\gamma} &  1 & 1 &  1\\
1 &  \gamma  & 1 &  1
\end{array}\right]=\left[\begin{array}{cccc}
1 &  1 & 1 &  1\\
1 &  1 & 1 &  1\\
1 &  1 & 1 &  1\\
1 &  1  & 1 &  1
\end{array}\right]+\left[\begin{array}{cccc}
0 &  0 & \tau &  0\\
0 &  0 & 0 &  \bar{\tau}\\
\bar{\tau} &  0 & 0 &  0\\
0 &  \tau  & 0 &  0
\end{array}\right],
\end{equation}
where $\tau=\gamma-1$. By using (\ref{tmp}) and Theorem 8.7 in
\cite{P} it is easy to show that
\begin{equation}\label{tmp-2}
x_k\doteq\|\Upsilon_{D(\theta_k)}-\id_4\|_{\mathrm{cb}}\leq|\tau_k|=|1-\gamma_k|=\left|1-\exp\!\left(\textstyle\frac{\mathrm{i}}{2}\theta_k\right)
\right|\leq\textstyle\frac{1}{2}|\theta_k|.
\end{equation}
Let
$\Delta_n\doteq\|\bigotimes_{k=1}^n\!\Upsilon_{D(\theta_k)}-\id_{4^n}\|_{\mathrm{cb}}$.
Then by using multiplicativity of the $cb$-norm and (\ref{tmp-2}) we
obtain
\begin{equation}\label{tmp-1}
\Delta_n\leq
x_{n}\prod_{k=1}^{n-1}(1+x_k)+\Delta_{n-1}\leq\prod_{k=1}^{n}(1+x_k)-1\leq\prod_{k=1}^{n}(1+\textstyle\frac{1}{2}|\theta_k|)-1.
\end{equation}
Assume that
$\,\bar{Q}_0\left(\bigotimes_{k=1}^n\!\Phi_{\theta_k}\right)>0$.
Then Lemma \ref{trans-l+} implies existence of unit vectors
$\varphi$ and $\psi$ in $\,\H^{\otimes n}_A=\mathbb{C}^{4^n}$ such
that
\begin{equation*}
\langle \psi|\Psi(A)|\varphi\rangle=0\quad\textup{and}\quad \langle
\varphi|\Psi(A)|\varphi\rangle=\langle
\psi|\Psi(A)|\psi\rangle\quad\forall A\in\mathfrak{L}_{0}^{\otimes
n},
\end{equation*}
where $\Psi=\bigotimes_{k=1}^n\!\Upsilon_{D(\theta_k)}$. Hence for
any $A$ in the unit ball of $\mathfrak{L}_{0}^{\otimes n}$ we have
$$
|\langle \psi|A|\varphi\rangle|\leq \Delta_n\quad\textup{and}\quad
|\langle \varphi|A|\varphi\rangle-\langle \psi|A|\psi\rangle|\leq
2\Delta_n
$$
By using (\ref{tmp-1}) and the inequality $x\geq\ln(1+x)$ it is easy
to see that the assumption $\,\sum_{k=1}^n|\theta_k|\leq
2\ln(3/2)\,$ implies $\Delta_n\leq1/2$. So, the above relations can
not be valid by the below Lemma \ref{b-lemma}, since
$\mathfrak{L}_{0}^{\otimes n}$ is a maximal commutative
$*$-subalgebra of $\mathfrak{M}_{4^n}$. $\square$\medskip

\begin{lemma}\label{b-lemma}
\emph{Let $\,\mathfrak{A}$ be a maximal commutative $*$-subalgebra of
$\,\mathfrak{M}_n$. Then
\begin{equation*}
\textit{either}\quad2\sup_{A\in\mathfrak{A}_1}\left|\langle
\psi|A|\varphi\rangle\right|>1\quad\textit{or}\quad
\sup_{A\in\mathfrak{A}_1}\left|\langle
|\varphi|A|\varphi\rangle-\langle \psi|A|\psi\rangle\right|>1
\end{equation*}
for any two unit vectors $\varphi$ and $\psi$ in $\mathbb{C}^n$,
where $\,\mathfrak{A}_1$ is the unit ball of $\,\mathfrak{A}$.}
\end{lemma}\smallskip

\begin{proof}
Let $\{x_i\}_{i=1}^n$ and $\{y_i\}_{i=1}^n$ be the coordinates of
$\varphi$ and $\psi$ in the basis in which the algebra
$\,\mathfrak{A}$ consists of diagonal matrices. Then
$$
\sup_{A\in\mathfrak{A}_1}\left|\langle
\psi|A|\varphi\rangle\right|=\sum_{i=1}^n|x_i||y_i|, \quad
\sup_{A\in\mathfrak{A}_1}\left|\langle
|\varphi|A|\varphi\rangle-\langle
\psi|A|\psi\rangle\right|=\sum_{i=1}^n\left||x_i|^2-|y_i|^2\right|.
$$
Let $d_i=|y_i|-|x_i|$. Assume that
$$
2\sum_{i=1}^n|x_i||y_i|\leq1\quad\text{and}\quad\sum_{i=1}^n\left||x_i|^2-|y_i|^2\right|\leq1.
$$
Since $\sum_{i=1}^n|x_i|^2=\sum_{i=1}^n|y_i|^2=1$, the first of
these inequalities implies
$$
\left|\sum_{i=1}^nd_i|x_i|\right|\geq 1/2\quad\text{and}\quad
\left|\sum_{i=1}^nd_i|y_i|\right|\geq 1/2.
$$
Hence
$$
\sum_{i=1}^n\left||x_i|^2-|y_i|^2\right|=\sum_{i=1}^n|d_i|[|x_i|+|y_i|]>\left|\sum_{i=1}^nd_i|x_i|\right|+\left|\sum_{i=1}^nd_i|y_i|\right|\geq1,
$$
where the strict inequality follows from the existence of negative
and positive numbers in the set $\{d_i\}_{i=1}^n$. This contradicts
to the above assumption.
\end{proof}

\section{Multi-dimensional generalization}

Note that
\begin{equation*}
\mathfrak{L}_0=\mathfrak{A}_{2}^{\otimes2},\quad \textrm{where}\quad
\mathfrak{A}_{2} = \left\{\left[\begin{array}{cc}
a &  b \\
b &  a
\end{array}\right]\!,\; a,b\in\mathbb{C}\,\right\}\!,
\end{equation*}
and that $\mathfrak{L}_{\theta}$ is the image of $\mathfrak{L}_0$
under the Schur multiplication by matrix (\ref{tmp}). So, the above
construction can be generalized by considering the corresponding
deformation of the maximal commutative $*$-subalgebra
$\mathfrak{L}^p_0=\mathfrak{A}_2^{\otimes p}$ of
$\mathfrak{M}_{2^p}$ for $p>2$. The algebra $\mathfrak{L}^p_0$ can
be described recursively as follows:
\begin{equation*}
\mathfrak{L}^p_0=\left\{\left[\begin{array}{cc}
A &  B \\
B &  A
\end{array}\right]\!,\; A,B\in\mathfrak{L}^{p-1}_0\right\},\quad
\mathfrak{L}^{1}_0=\mathfrak{A}_2.
\end{equation*}

Let $p>2$ and $\theta\in\mathrm{T}\doteq(-\pi,\pi]$ be arbitrary,
$\gamma=\exp\left(\frac{\mathrm{i}}{2}\theta\right)$. Let
$D(\theta)$ be the $2^{p}\times 2^{p}$ matrix described as
$2^{p-1}\times 2^{p-1}$ matrix $[A_{ij}]$ consisting of the blocks
$$
A_{ii}=\!\left[\begin{array}{cc}
1 &  1 \\
1 &  1
\end{array}\right]\;\forall i,\;\, A_{ij}=\!\left[\begin{array}{cc}
\gamma &  1 \\
1 &  \bar{\gamma}
\end{array}\right]\;\textrm{if}\;\, i<j\;\;\textrm{and}\;\, A_{ij}=\!\left[\begin{array}{cc}
\bar{\gamma} &  1 \\
1 & \gamma
\end{array}\right]\;\textrm{if}\;\, i>j.
$$

Consider the $2^{p}$-D subspace
$\mathfrak{L}^p_{\theta}=\Upsilon_{D(\theta)}\left(\mathfrak{L}^p_0\right)$
of $\mathfrak{M}_{2^p}$ ($\Upsilon_{D(\theta)}$ is the Schur
multiplication by the matrix $D(\theta)$).  This subspace satisfies
condition (\ref{L-cond}) and has the following property
\begin{equation}\label{W-inv+}
    A=W_{2^p}^*AW_{2^p}\quad \forall A\in \mathfrak{L}^p_{\theta},
\end{equation}
where $W_{2^p}$ is the $2^{p}\times 2^{p}$ matrix having $"1"$ on
the main skew-diagonal and $"0"$ on the other places. To prove
(\ref{W-inv+}) it suffices to show that it holds for the algebra
$\mathfrak{L}^p_{0}=\mathfrak{A}_2^{\otimes p}$ (by using
$W_{2^p}=W_{2}^{\otimes p}$) and to note that the map
$\Upsilon_{D(\theta)}$ commutes with the transformation $A\mapsto
W_{2^p}^*AW_{2^p}$.

Denote by $\widehat{\mathfrak{L}}^p_{\theta}$ the set of all
channels whose noncommutative graph coincides with
$\mathfrak{L}^p_{\theta}$. By Proposition 2 in \cite{Sh&Sh} the set
$\,\widehat{\mathfrak{L}}^p_{\theta}$ contains pseudo-diagonal
channels with $d_A=2^p$ and  $d_E$ such that $d_E^{\,2}\geq2^{p}$.
\smallskip

\begin{theorem}\label{sqc-2} \emph{Let $\,p>1$ and $\,n>1$ be given natural numbers, $\,\Phi_{\theta}$ be an arbitrary channel in
$\,\widehat{\mathfrak{L}}^p_{\theta}\,$  and
$\,\delta_p=\frac{1}{2^{p-1}}\sum\limits_{k=1}^{2^{p-1}}\left|\cot\left(\frac{(2k-1)\pi}{2^p}\right)\right|>0$.}\smallskip

 A) \emph{$\;\bar{Q}_{0}(\Phi^{\otimes n}_{\theta})=0\,$ if
$\;|\theta|\leq\theta_n$, where $\theta_n$ is the minimal positive
solution of the equation}
\begin{equation}\label{th-eq}
2(1-\cos(\theta/2))+\delta_p\sin(\theta/2)=n^{-1}\ln(3/2).
\end{equation}

B) \emph{If $\;\theta=\pm\pi/n$ then $\;\bar{Q}_{0}(\Phi^{\otimes
n}_{\theta})\geq p-1\,$  and  there exist $\,2^n$ mutually
orthogonal $\,2^{p-1}\textrm{-}\,$D error correcting codes for the
channel $\,\Phi^{\otimes n}_{\theta}$.  For each binary $n$-tuple
$(x_1,\ldots x_n)$ the corresponding error correcting code is
spanned by the image of the vectors
\begin{equation}\label{main-vec+}
|\varphi_k\rangle=\textstyle{\frac{1}{\sqrt{2}}}\left[\;|2k-1\ldots
2k-1\rangle+\mathrm{i}\,|2k\ldots 2k\rangle\,\right],\quad
k=\overline{1,2^{p-1}},
\end{equation}
under the unitary transformation $\,U_{x_1}\otimes\ldots\otimes
U_{x_n}$, where $\{|k\rangle\}$ is the canonical basis in
$\mathbb{C}^{2^p}$, $\,U_0=I_{2^p}$ and $\,U_1=W_{2^p}$ (defined in
(\ref{W-inv+})).}
\end{theorem}
\medskip

\begin{remark}\label{sqc-2r}
The constant $\delta_p$ is the Schur multiplier norm of the
skew-symmetric $2^{p-1}\times2^{p-1}$ matrix having $"1"$ everywhere
below the main diagonal. So, the sequence $\{\delta_p\}$ is
non-decreasing. It is easy to see that $\delta_2=1$,
$\delta_3=\sqrt{2}$, $\delta_4\approx1.84$ and that
$\delta_p=\left(\frac{2\ln2}{\pi}\right)p+o(p)$ for large $p$
\cite{M}.\smallskip

Note also that
$\theta_n=2\ln(3/2)\left(n\delta_p\right)^{-1}+o(1/n)$ for large
$n$.
\end{remark}
\medskip
\begin{remark}\label{sqc-2rr} Assertion B of Theorem 2 can be
strengthened as follows:

B') \emph{If $\;\theta=\pm\pi/n\,$ then there exist $\,2^n$ mutually
orthogonal $\,2^{p-1}\textrm{-}\,$D projectors $P_{\bar{x}}$ indexed
by a binary $n$-tuple $\,\bar{x}=(x_1,\ldots x_n)$ such that
$$
P_{\bar{x}}AP_{\bar{x}}=\lambda(A)P_{\bar{x}}\quad \forall
A\in[\mathfrak{L}^p_{\theta}]^{\otimes n},
$$
where $\lambda(A)\in\mathbb{C}$ does not depend on $\,\bar{x}$.
$P_{\bar{x}}$ is the projector on the subspace
$\,U_{x_1}\otimes\ldots\otimes U_{x_n}(\H_0)$, where $\H_0$ is the
linear hull of vectors (\ref{main-vec+}).}

So, in the orthonormal basis $\,\left\{U_{x_1}\otimes\ldots\otimes
U_{x_n}|\varphi_1\rangle, U_{x_1}\otimes\ldots\otimes
U_{x_n}|\varphi_2\rangle \ldots \right\}$ the main
$\,2^{n+1}\times2^{n+1}$ minor of all matrices in
$[\mathfrak{L}^p_{\theta}]^{\otimes n}$ has form (\ref{min-f}) with
$I_2$ replaced by $I_{2^{p-1}}$.
\end{remark}\smallskip

\emph{Proof of Theorem 2.} A)  Note that
$\,\Upsilon_{D(\theta)}-\id_{2^p}\,$ is the Schur multiplication by
the matrix
$$
-T\otimes\left[\begin{array}{cc}
u &  0 \\
0 &  u
\end{array}\right]+S\otimes\left[\begin{array}{cc}
\bar{v} &  0 \\
0 & v
\end{array}\right],
$$
where $T$ is the $2^{p-1}\times 2^{p-1}$ matrix having $"0"$ on the
main diagonal and $"1"$ on the other places, $S$ is the
$2^{p-1}\times 2^{p-1}$ skew-symmetric matrix having $"1"$
everywhere below the main diagonal,
$u=1-\Re\gamma=1-\cos[\theta/2],\;
v=\mathrm{i}\Im\gamma=\mathrm{i}\sin[\theta/2]$.

In \cite{M} it is shown that
$\|\Upsilon_{S}\|_{\mathrm{cb}}=2^{1-p}\|S\|_1=\delta_p$. Since
$\|\Upsilon_{T}\|_{\mathrm{cb}}\leq2$ and $\|\Upsilon_{A\otimes
B}\|_{\mathrm{cb}}=\|\Upsilon_{A}\otimes\Upsilon_{B}\|_{\mathrm{cb}}=\|\Upsilon_{A}\|_{\mathrm{cb}}\|\Upsilon_{B}\|_{\mathrm{cb}}$,
we have
$$
x\doteq\|\Upsilon_{D(\theta)}-\id_{2^p}\|_{\mathrm{cb}}\leq
u\|\Upsilon_{T}\|_{\mathrm{cb}}+|v|\|\Upsilon_{S}\|_{\mathrm{cb}}=
2(1-\cos(\theta/2))+\delta_p|\sin(\theta/2)|
$$
and hence $x\leq n^{-1}\ln(3/2)\leq\sqrt[n]{3/2}-1$ if
$|\theta|\leq\theta_n$.\medskip

Assume that $\bar{Q}_{0}(\Phi^{\otimes n}_{\theta})>0\,$ for some
$\theta\in[-\theta_n, \theta_n]$.  By repeating the arguments from
the proof of part C of Theorem \ref{sqc} we obtain
\begin{equation}\label{as-rel}
|\langle \psi|A|\varphi\rangle|\leq \Delta_n\quad\textup{and}\quad
|\langle \varphi|A|\varphi\rangle-\langle \psi|A|\psi\rangle|\leq
2\Delta_n
\end{equation}
for some unit vectors $\varphi,\psi\in\mathbb{C}^{2pn}$ and all $A$
in the unit ball of $[\mathfrak{L}^p_{0}]^{\otimes n}$, where
$$
\Delta_n\doteq\|\Upsilon_{D(\theta)}^{\otimes
n}-\id_{2^{pn}}\|_{\mathrm{cb}}\leq (x+1)^n-1\leq1/2.
$$
Since $[\mathfrak{L}^p_{0}]^{\otimes n}$ is a maximal commutative
$*$-subalgebra of $\mathfrak{M}_{2^{pn}}$, Lemma \ref{b-lemma} shows
that (\ref{as-rel}) can not be valid.\medskip

B) Let $\theta=\pm\pi/n$. To prove that the linear hull $\H_0$ of
vectors (\ref{main-vec+}) is an error correcting code for the
channel $\mathrm{\Phi}^{\otimes n}_{\theta}$ it suffices, by Lemma
\ref{trans-l+}, to show that
\begin{equation*}
    \langle\varphi_l | M_1\otimes \ldots\otimes M_n | \varphi_k\rangle=0\quad \forall
    M_1,\ldots,M_n\in\mathfrak{L}^p_{\theta},\; \forall
    k,l
\end{equation*}
and that
\begin{equation*}
    \langle\varphi_l | M_1\otimes \ldots\otimes M_n |\varphi_l\rangle=\langle\varphi_k | M_1\otimes \ldots\otimes M_n | \varphi_k\rangle\quad \forall
    M_1,\ldots,M_n\in\mathfrak{L}^p_{\theta},\; \forall
    k,l.
\end{equation*}
Since any matrix in $\mathfrak{L}^p_{\theta}$ can be described as
$2^{p-1}\times 2^{p-1}$ matrix $[A_{ij}]$ consisting of the blocks
$$
A_{ii}=\!\left[\begin{array}{cc}
a &  b \\
b &  a
\end{array}\right]\;\forall i\quad\textrm{and}\quad A_{ij}=\!\left[\begin{array}{cc}
\bar{\gamma}_{ij} c_{ij} &  d_{ij} \\
d_{ij} &  \gamma_{ij} c_{ij}
\end{array}\right]\quad\forall i\neq j,
$$
where $\,\gamma_{ij}=\exp\left(\mathrm{i}s_{ij}\theta/2\right)$,
$\,s_{ij}=\mathrm{sgn}(j-i)\,$ and $\,a,b,c_{ij},d_{ij}$ are some
complex numbers, the above relations are proved by the same way as
in the proof of part B of Theorem \ref{sqc} (by  using
$\gamma^n_{ij}+\bar{\gamma}^n_{ij}=0$).

To prove that the subspace $U_{\bar{x}}(\H_0)$, where
$\,U_{\bar{x}}=U_{x_1}\otimes\ldots\otimes U_{x_n}$, is an error
correcting code  for the channel $\mathrm{\Phi}^{\otimes
n}_{\theta}$ it suffices to note that (\ref{W-inv+}) implies
$U^*_{\bar{x}}AU_{\bar{x}}=A$ for all
$A\in[\mathfrak{L}^p_{\theta}]^{\otimes n}$. $\square$\medskip

\begin{corollary}\label{sqc-2c} \emph{Let $\;n\,$ be arbitrary
and $\;m\,$ be a natural number such that
$\theta_*=\pi/m\leq\theta_n$. Then}
\begin{equation*}
\bar{Q}_0(\Phi_{\theta_*}^{\otimes n})=0\;\;\;
\textit{but}\;\;\;\bar{Q}_0(\Phi_{\theta_*}^{\otimes m})\geq
p-1\;\;\;\textit{and hence}\;\;\; Q_0(\Phi_{\theta_*})\geq (p-1)/m.
\end{equation*}
\emph{There exist $\,2^m$ mutually orthogonal $\,2^{p-1}$-D error
correcting codes for the channel $\,\Phi_{\theta_*}^{\otimes m}$.}
\end{corollary}
\medskip

\begin{remark}\label{sqc-2r+}
Corollary \ref{sqc-2c} (with Proposition 2 in \cite{Sh&Sh} and
Remark \ref{sqc-2r}) shows that for any $n$ there exists a channel
$\Phi_n$ with $d_A=2^p$ and arbitrary $d_E$ satisfying the
inequality $d_E^{\,2}\geq2^{p}$ such that
$$
\bar{Q}_0(\Phi_n^{\otimes n})=0\quad \textrm{and} \quad
Q_0(\Phi_n)\geq
\frac{p-1}{[\pi/\theta_n]+1}=\frac{2\ln(3/2)(p-1)}{\pi n
\delta_p}+o(1/n),
$$
where $[x]$ is the integer part of $x$, and hence we have the
following lower bounds for the values $S_{d}(n)$ and $S_{*}(n)$
(introduced in (\ref{s-d-def}) and (\ref{s-star-def}))
$$
S_{2^p}\geq\frac{2\ln(3/2)(p-1)}{\pi n
\delta_p}+o(1/n)\quad\textrm{and}\quad S_{*}(n)\geq
\frac{2\ln(3/2)(p-1)}{\pi n \delta_p}
$$
(the later inequality is obtained from the former by using relation
(\ref{s-rel})).
\end{remark}
\medskip

Since $\delta_2=1$, the above lower bounds with $p=2$ coincide with
(\ref{s-e-0})-(\ref{s-e-2}).


Since $\delta_3=\sqrt{2}$, Remark \ref{sqc-2r+} with $\,p=3\,$ shows
that for any $n$ there exists a channel $\Phi_n$ with $d_A=8$ and
$d_E=3$ such that
$$
\bar{Q}_0(\Phi_n^{\otimes n})=0\quad \textrm{and} \quad
Q_0(\Phi_n)\geq \sqrt{2}\times\frac{2\ln(3/2)}{\pi n}+o(1/n).
$$
Hence
$$
S_8(n)\geq\sqrt{2}\times\frac{2\ln(3/2)}{\pi n}+o(1/n).
$$
Comparing this estimation with (\ref{s-e-0}), we see that the
increasing input dimension $d_A$ from $\,4\,$ to $\,8\,$ gives the
amplification factor $\sqrt{2}$ for the quantum zero-error capacity
of a channel having vanishing $n$-shot capacity (more precisely, for
the lower bound of this capacity).
\medskip

In general, Remark \ref{sqc-2r+} shows that our construction with
the input dimension $d_A=2^p$  amplifies lower bound (\ref{s-e-2})
for $S_*(n)$ by the factor $\Lambda_p=\frac{p-1}{\delta_p}$. By
Remark \ref{sqc-2r} the non-decreasing sequence $\Lambda_p$ has a
finite limit:
$$
\lim_{p\rightarrow+\infty}\Lambda_p=\Lambda_*\doteq\frac{\pi}{2\ln
2}\approx 2.26.
$$
Hence  $\,\Lambda_*\approx 2.26\,$ is the maximal amplification
factor for $S_*(n)$ which can be obtained by increasing input
dimension. So, we have
$$
S_*(n)\geq \Lambda_*\frac{2\ln(3/2)}{\pi n
}=\frac{\log_2(3/2)}{n}\qquad \forall n.
$$
Unfortunately, we have not managed to show existence of a channel
with \emph{arbitrary}  quantum zero-error capacity and vanishing
$n$-shot capacity, i.e. to prove the conjecture $S_*(n)=+\infty$ for
all $n$. This can be explained as follows.

According to Theorem \ref{sqc-2}, if the input dimension of the
channel $\Phi_{\theta}$ increases as $2^p$ then the dimension of
error-correcting code for the channel $\Phi^{\otimes m}_{\theta}$,
$\theta=\pi/m$, increases as $2^{p-1}$. But simultaneously the norm
of the map $\,\Upsilon_{D(\theta)}-\id_{2^p}\,$ characterizing
deformation of a maximal commutative
$*$\nobreakdash-\hspace{0pt}subalgebra increases as
$\,\delta_p\sin(\theta/2)\sim p\theta/2\,$ for large $p$ and small
$\theta$, so, to guarantee vanishing of the $n$-shot capacity of
$\Phi_{\theta}$ by using Lemma \ref{b-lemma} we have to decrease the
value of $\theta$ as $O(1/p)$. Since $\theta=\pi/m$, we see that
$\bar{Q}_0(\Phi^{\otimes m}_{\theta})$ and $m$ have the same
increasing rate $O(p)$, which does not allow to obtain large values
of $Q_0(\Phi_{\theta})$.
\smallskip

Thus, the main obstacle for proving the conjecture $S_*(n)=+\infty$
consists in the unavoidable growth of the norm of the map
$\,\Upsilon_{D(\theta)}-\id_{2^p}\,$  as $\,p\rightarrow+\infty$
(for fixed $\theta$).

First there was a hope to solve this problem by using a freedom in
choice of the deformation map $\mathrm{\Upsilon}_{D(\theta)}$.
Indeed, instead of the matrix $D(\theta)$ introduced before the
definition of $\mathfrak{L}^p_{\theta}$ one can use the  matrix
$D(\theta,S)=[A_{ij}]$ consisting of the blocks
$$
A_{ii}=\!\left[\begin{array}{cc}
1 &  1 \\
1 &  1
\end{array}\right]\;\forall i,\; A_{ij}=\!\left[\begin{array}{cc}
\gamma &  1 \\
1 &  \bar{\gamma}
\end{array}\right]\;\textrm{if}\; s_{ij}=-1\;\,\textrm{and}\; A_{ij}=\!\left[\begin{array}{cc}
\bar{\gamma} &  1 \\
1 & \gamma
\end{array}\right]\;\textrm{if}\; s_{ij}=1,
$$
where $S=[s_{ij}]$ is \emph{any} skew-symmetric $2^{p-1}\times
2^{p-1}$ matrix such that $s_{ij}=\pm1$ for all $i\neq j$. For the
corresponding subspace
$\mathfrak{L}^p_{\theta,S}=\mathrm{\Upsilon}_{D(\theta,S)}\left(\mathfrak{L}^p_0\right)$
the main assertions of Theorem 2 are valid (excepting the assertion
about $2^m$ error correcting codes) with the constant $\delta_p$
replaced by the norm $\|\mathrm{\Upsilon}_{S}\|_{\mathrm{cb}}$ (in
our construction $S=S_*$ is the matrix having $"1"$ everywhere below
the main diagonal and
$\delta_p=\|\mathrm{\Upsilon}_{S_*}\|_{\mathrm{cb}}$). But the
further analysis (based on the results from \cite{M}) has shown that
$$
\|\mathrm{\Upsilon}_{S}\|_{\mathrm{cb}}\geq\delta_p=\|\mathrm{\Upsilon}_{S_*}\|_{\mathrm{cb}}
$$
and hence
$$
\|\mathrm{\Upsilon}_{D(\theta,S)}-\id_{2^p}\,\|_{\mathrm{cb}}\geq
\|\mathrm{\Upsilon}_{D(\theta,S_*)}-\id_{2^p}\,\|_{\mathrm{cb}}
$$
for any skew-symmetric $2^{p-1}\times 2^{p-1}$ matrix $S$ such that
$s_{ij}=\pm1$ for all $i\neq j$. So, by using the above modification
we can not increase the lower bound for
$Q_0(\mathrm{\Phi}_{\theta})$. The useless of some other
modifications of the map $\mathrm{\Upsilon}_{D(\theta)}$ was also
shown.

It is interesting to note that the norm growth  of the map
$\,\mathrm{\Upsilon}_{D(\theta)}-\id_{2^p}\,$ is a \emph{cost of the
symmetry requirement} for the subspace $\mathfrak{L}^p_{\theta}$.
Indeed, if we omit this requirement then we would use the matrix
$\widetilde{D}(\theta)=[A_{ij}]$ consisting of the blocks
$$
A_{ii}=\left[\begin{array}{cc}
1 &  1 \\
1 &  1
\end{array}\right]\;\forall i\quad\textrm{and}\quad A_{ij}=\left[\begin{array}{cc}
\gamma &  1 \\
1 &  \bar{\gamma}
\end{array}\right]\;\forall i\neq j,
$$
for which
$\,\|\mathrm{\Upsilon}_{\widetilde{D}(\theta)}-\id_{2^p}\|_{\mathrm{cb}}\leq
2|\gamma-1|\leq\theta\,$ for all $p$.

It seems that the above obstacle is technical and can be overcome
(within the same construction of a channel) by finding a way to
prove the equality $\bar{Q}_0(\mathrm{\Phi}^{\otimes n})=0\,$ not
using estimations of the distance between the unit balls of
$[\mathfrak{L}^p_{\theta}]^{\otimes n}$ and of
$[\mathfrak{L}^0_{\theta}]^{\otimes n}$. Anyway the question
concerning the value
\begin{equation*}
S_*(n)\doteq\sup_{\mathrm{\Phi}}\left\{Q_0(\mathrm{\Phi})\,|\,\bar{Q}_0(\mathrm{\Phi}^{\otimes
n})=0\,\right\}
\end{equation*}
remains open.

\section*{Appendix: Stinespring representations for the channels $\Phi^1_{\theta}$ and $\Phi^2_{\theta}$}

\emph{Proof of Lemma \ref{cfc}}. Show first that for each $\theta$
one can construct basis $\{A^{\theta}_i\}_{i=1}^4$ of
$\mathfrak{L}_{\theta}$ consisting of positive operators with
$\,\sum_{i=1}^4 A^{\theta}_i=I_4$ such that:
\begin{enumerate}[1)]
    \item the
function $\theta\mapsto A^{\theta}_i$ is continuous for
$i=\overline{1,4}$;
    \item $\{A^{0}_i\}_{i=1}^4$ consists of mutually orthogonal 1-rank
    projectors.
\end{enumerate}

Note that $\mathfrak{L}_{\theta}$ is unitary equivalent to the
subspace $\mathfrak{L}^s_{\theta}$ defined by (\ref{L-copy}).

Denote by $\|T_{\theta}\|$ the operator norm of the matrix
$T_{\theta}$ involved in (\ref{L-copy}). Note that the function
$\theta\mapsto T_{\theta}$ is continuous, $T_0=0$ and
$\|T_{\theta}\|\leq\|T_{\pi}\|=2$. Let
\begin{equation*}
\tilde{A}^{\theta}_1=\left[\begin{array}{cccc}
\alpha &  0 & 0 &  0\\
0 &  \beta & 0 &  0\\
0 &  0 & \beta &  0\\
0 &  0 & 0 &  \beta
\end{array}\right]-\textstyle\frac{1}{4}(\alpha-\beta)\,T_{\theta},\quad \tilde{A}^{\theta}_2=\left[\begin{array}{cccc}
\beta &  0 & 0 &  0\\
0 &  \alpha & 0 &  0\\
0 &  0 & \beta &  0\\
0 &  0 & 0 &  \beta
\end{array}\right]-\textstyle\frac{1}{4}(\alpha-\beta)\,T_{\theta},
\end{equation*}
\begin{equation*}
\tilde{A}^{\theta}_3=\left[\begin{array}{cccc}
\beta &  0 & 0 &  0\\
0 &  \beta & 0 &  0\\
0 &  0 & \alpha &  0\\
0 &  0 & 0 &  \beta
\end{array}\right]+\textstyle\frac{1}{4}(\alpha-\beta)\,T_{\theta},\quad \tilde{A}^{\theta}_4=\left[\begin{array}{cccc}
\beta &  0 & 0 &  0\\
0 &  \beta & 0 &  0\\
0 &  0 & \beta &  0\\
0 &  0 & 0 &  \alpha
\end{array}\right]+\textstyle\frac{1}{4}(\alpha-\beta)\,T_{\theta}
\end{equation*}
be operators in $\mathfrak{L}^s_{\theta}$, where
$\beta=\min\left\{\frac{3}{16},\frac{1}{4}\|T_{\theta}\|\right\}$
and $\alpha=1-3\beta$. It is easy to verify that
$\,\tilde{A}^{\theta}_i\geq0$ for all $i$ and $\,\sum_{i=1}^4
\tilde{A}^{\theta}_i=I_4$.  Then
$\{A^{\theta}_i=S\tilde{A}^{\theta}_iS^{-1}\}_{i=1}^4$, where $S$ is
the unitary matrix defined before (\ref{L-copy}), is a required
basis of $\mathfrak{L}_{\theta}$.

Let $m\geq2$ and $\,\{|\psi_i\rangle\}_{i=1}^4$ be a collection of
unit vectors in $\mathbb{C}^m$ such that
$\,\{|\psi_i\rangle\langle\psi_i|\}_{i=1}^4$ is a linearly
independent subset of $\,\mathfrak{M}_m$. It is easy to show (see
the proof of Corollary 1 in \cite{Sh&Sh}) that
$\mathfrak{L}_{\theta}$ is a noncommutative graph of the
pseudo-diagonal channel
$$
\Phi_{\theta}(\rho)=\Tr_{\mathbb{C}^m}V_{\theta}\rho V_{\theta}^*,
$$
where
$$
V_{\theta}:|\varphi\rangle\mapsto\sum_{i=1}^4[A^{\theta}_i]^{1/2}|\varphi\rangle\otimes|i\rangle\otimes|\psi_i\rangle
$$
is an isometry from $\H_A=\mathbb{C}^4$ into
$\mathbb{C}^4\otimes\mathbb{C}^4\otimes\mathbb{C}^m$
($\{|i\rangle\}$ is the canonical basis in $\mathbb{C}^4$). By
property 1 of the basis $\{A^{\theta}_i\}_{i=1}^4$ the function
$\theta\mapsto V_{\theta}$ is continuous.\smallskip

The first part of Lemma \ref{cfc} follows from this construction
with $m=2$.

To prove the second part assume that $m=4$ and
$|\psi_i\rangle=|i\rangle$, $i=\overline{1,4}$. Property 2 of the
basis $\{A^{\theta}_i\}_{i=1}^4$  implies
$$
V_{0}|\varphi\rangle=\sum_{i=1}^4\langle
e_i|\varphi\rangle|e_i\rangle\otimes|i\rangle\otimes|i\rangle,
$$
where $\,\{|e_i\rangle\}_{i=1}^4$ is an orthonormal basis in
$\mathbb{C}^4$. Hence $\Phi_{0}(\rho)=\sum_{i=1}^4\langle
e_i|\rho|e_i\rangle\sigma_i$, $\sigma_i=|e_i\otimes i\rangle\langle
e_i\otimes i|$, is a classical-quantum channel.

\bigskip

I am grateful to A.S.Holevo and to the participants of his seminar
"Quantum probability, statistic, information" (the Steklov
Mathematical Institute) for useful discussion. I am also grateful to
T.Shulman and P.Yaskov for the help in solving the particular
questions.

\medskip

Note Added: After publication of the first version of this
paper the analogous result concerning quantum $\varepsilon$-error
capacity has been appeared \cite{C&Co}.

\end{document}